\documentclass[useAMS,usegraphicx]{mn2e}
\usepackage{color}

\def\lesssim{\,\lower2truept\hbox{${<\atop\hbox{\raise4truept\hbox{$\sim$}}}$}\,}
\def\gtrsim{\,\lower2truept\hbox{${>\atop\hbox{\raise4truept\hbox{$\sim$}}}$}\,}

\def\rev{}

\title[Mass Loss and DMH Profiles]{Effects of Baryon Mass Loss on Profiles of
Large Galactic Dark Matter Haloes}

\author[Ragone-Figueroa, Granato \& Abadi]{
\parbox[t]{\textwidth}{
Cinthia Ragone-Figueroa$^1$\thanks{Email: cin@oac.uncor.edu},
Gian~Luigi Granato$^{2}$\thanks{Email: granato@oats.inaf.it} \&
Mario G.\ Abadi$^1$\thanks{Email: mario@oac.uncor.edu}
}
\vspace*{6pt} \\
  $^1$ Instituto de Astronom\'ia Te\'orica y Experimental, IATE,
  CONICET-Observatorio
  Astron\'omico, \\
  Universidad Nacional de C\'ordoba, Laprida 854, X5000BGR, C\'ordoba,
  Argentina\\
  $^2$ Istituto Nazionale di Astrofisica INAF, Osservatorio Astronomico di
  Trieste, Via
  Tiepolo
  11, I-34131
  Trieste, Italy \\
}

\begin{document}
\newcommand{\cinthia}[1]{{\bf\textcolor{red}{ #1}}}
\date{Accepted ... Received ...}

\maketitle

\begin{abstract}
We perform controlled numerical experiments to asses the effect of baryon mass
loss on the inner structure of large galactic dark matter haloes. This mass
expulsion is intended to mimic both the supernovae and AGN feedbacks, as well
as the evolution of stellar populations. {\rev This study is meant in
particular for precursors of massive Early Type Galaxies, wherein strong AGN
feedback (often dubbed "QSO mode" in galaxy formation models)} has been
proposed to remove on a short timescale, of the order of a few dynamical times,
a substantial fraction of their baryons. In a previous paper we evaluated the
observational consequences (size increase) of this process on the galactic
structure (Ragone-Figueroa \& Granato 2011). Here we focus on the distribution
of dark matter in the galactic region. It is shown that the inner region of the
DM halo expands and its density profile flattens by a sizeable amount, with
little dependence on the expulsion timescale. We also evaluate the effect of
the commonly made approximation of treating the baryonic component as a
potential that changes in intensity without any variation in shape. This
approximation leads to some underestimates of the halo expansion and its slope
flattening. We conclude that cuspy density profiles in ETGs could be difficult
to reconcile with an effective AGN (or stellar) feedback during the evolution
of these systems.

\end{abstract}


\section{Introduction}
\label{sec:intro} A long standing puzzle of post recombination cosmology based
on Cold Dark Matter, independently of the presence of cosmological constant or
curvature in the adopted cosmological model, is the so called {\it core-cusp}
problem (for a recent review see De Blok 2010). Since the 90's, a long series
of N-body (gravity only) simulations, dating back to Dubinsky \& Carlberg
(1991), has produced Dark Matter Haloes (DMH) whose inner density profile is
reasonably well described by a power law $\rho \propto r^{\alpha}$ with $\alpha
\sim -1$, i.e.\ a cusp. This is at odd with several observations, suggesting a
flat density profiles of DM in the inner region of real galaxies, i.e.\ a core.
The case for cored DM density profiles is rather strong in dwarf and disc
dominated Low Surface Brightness (LSB) galaxies, wherein the dynamics of
visible matter tracks safely the largely dominating gravitational field of DM.
Also the dynamics of normal spiral galaxies is best interpreted by means of
cored DM mass models (e.g.\ Salucci \& Frigerio-Martins 2009 and references
therein), while the presence of a core in DMH of large elliptical galaxies is
difficult to assess and controversial (e.g.\ Memola, Salucci \& Babi{\'c} 2011;
Sonnenfeld et al.\ 2011; Tortora et al.\ 2011), since their central region is
even more gravitationally dominated by the baryon component, and because of the
less straight-forward to interpret orbital structure.

A widespread idea is that the solution of the core-cusp problem should be
searched for in the physics of baryonic matter, that can affect to some extent
the distribution of DM in the inner region of haloes, and is not included in
the aforementioned computations.

However, the first baryonic process expected to occur worsen the problem. It
consists in some further contraction of the central region of the DMH, induced
by the much stronger concentration-collapse of dissipative baryonic matter
(e.g.\ Blumenthal et al.\ 1986; Gnedin et al.\ 2004; Abadi et al.\ 2010, Gnedin
et al.\ 2011). The exact importance of this process is somewhat debated.
Different numerical works found different results, and significantly less
contraction than that predicted by the simple analytic estimate by Blumenthal
et al.\ (1986). In any case, this mechanism produces some further steepening of
the inner profile, with respect to the prediction of gravity only cosmological
simulations.

Nevertheless, several subsequent (and more complex) baryon processes occur
after this primary condensation, and can act in the opposite direction. Indeed,
they have been often evaluated specifically to seek solutions to the core-cusp
problem (e.g.\ Navarro, Eke \& Frenk 1996; Gnedin \& Zhao 2002; Read \& Gilmore
2005; Tonini, Lapi \& Salucci 2006; Mashchenko, Couchman \& Wadsley 2006;
Mashchenko, Wadsley \& Couchman 2008; Pasetto et al.\ 2010; Governato et al.\
2010; De Souza et al.\ 2011; Ogiya \& Mori 2011; Inoue \& Saitoh 2011; {\rev
Martizzi et al.\ 2011; Pontzen \& Governato 2012; Macci{\`o} et al.\ 2012}).
Several of these papers have addressed the effect of baryon ejection, due to
SNae feedback, on the central DM density profile of less massive systems, i.e.\
mainly dwarfs or at most disk dominated galaxies. The results on its importance
have been sometimes contradictory (for a discussion of possible causes, see
Read \& Gilmore 2005).

Much less attention has been paid so far to these effects in larger systems,
such as Early Type Galaxies (ETGs). The main reason is obviously that their DM
content is less constrained by the data, as noted above (for a review see Buote
\& Humphrey 2012). Moreover, mechanisms related to SNae feedback are expected
to be less effective in the deeper potential wells of large ETGs. As for the
first point, it is worth noticing that recent observational evidence points
either to a DM distribution shallower (Memola, Salucci \& Babi{\'c} 2011) or
cuspier (Sonnenfeld et al. 2011; Tortora et al.\ 2012) than that predicted by
$\Lambda$CDM gravity only simulations. New techniques could soon provide
constrains on the DM distribution in ellipticals (e.g.\ Pooley et al.\ 2011).
As for the second point, in the past few years it has become common to consider
in galaxy formation theory the feedback of AGN activity, which is likely to be
much more effective in massive systems than that due to SNae (e.g.\ Silk \&
Rees 1998; Fabian 1999; Granato et al. 2001, 2004; Benson et al.\ 2003;
Cattaneo et al. 2006; Monaco et al. 2007; Sijacki et al. 2007; Somerville et
al. 2008; Johansson et al. 2009; Ciotti, Ostriker \& Proga 2009). AGN feedback
is widely considered the most promising mechanism for relaxing those tensions
between galaxy formation models and observation, broadly related to the so
called {\it overcooling problem}. This is the tendency of models to produce too
massive and too blue (i.e.\ star forming) galaxies at low redshift, and to lock
up in galaxies an excessive fraction of available baryons. On the contrary,
many observations indicate that on average the more massive a galaxy, the
earlier it stops its star formation activity, a phenomenon referred to as {\it
downsizing} (e.g. Brammer et al.\ 2011; for a critical discussion of the
various manifestations of downsizing see Fontanot et al. 2009). Despite the
likely prominent role of AGN feedback in galaxy formation, so far its effect on
the DM distribution has not been evaluated in detail (but see Peirani, Kay \&
Silk 2008 and Duffy et al.\ 2010).

One kind of AGN feedback (that sometimes called "QSO mode") is expected to
eject on a small timescale (of the order of a few 10 Myr at most) the cold gas
not yet converted into stars in star forming ETGs. In a recent work
(Ragone-Figueroa \& Granato 2011; henceforth Paper I) we have evaluated with
aimed numerical simulations the importance of this process, concentrating our
analysis on the baryon component of ETGs, in order to asses its possible
contribution to the observed size evolution of ETGs (see e.g.\ Newmann et al.\
2011 and references therein). In this paper we investigate instead in detail
the effects on the profile of the DM halo.

Since we are mainly interested in ETGs rather than less massive systems such as
dwarfs, we will extend the studies already published to regions of the
parameter space not covered previously. Indeed, this is the first work where
the initial conditions have been thought to get a configuration, after the loss
of a substantial fraction of baryons previously condensed in the central region
of the DM halo (i.e.\ the galactic region), consistent with our basic knowledge
of the properties of local large ETGs (baryon to DM mass ratio, scalelengths,
size as a function of stellar mass). By converse, in most studies the baryons
totally disappear, or the remaining fraction is $\le 5\%$, consistently with
the very low baryon content of dwarf galaxies. This is not the case for ETGs,
where the leftover baryons still dominate the potential wells in the central
region of the DM halo. Also, in order to asses the maximal effect of baryon
loss, in most studies the initial mass ratio of condensed baryon to DM has been
set close to the cosmic baryon fraction. {\rev However this appears too
extreme, at least for ETGs. According to the cooling prescriptions adopted by
semi analytic models and simulations of galaxy formation,  no more than a few
tens percent of the cosmic baryons had time to cool and condense in a large
galactic DM halo, during the few Gyrs over which ETGs formed at $z\gtrsim 1.5$
(e.g.\ Benson et al.\ 2001; Helly et al.\ 2003;  Granato et al.\ 2004; Cattaneo
et al.\ 2007; Viola et al.\ 2008).} {\rev We point out explicitly that the
results of previous numerical experiments, when featuring substantial
differences in the initial and/or final mass ratios of baryon to DM, or in the
assumed density profiles of the two components (often disks for the baryons),
cannot be simply re-scaled to predict quantitatively our findings. }

Moreover, {\rev a technical difference with respect to} most previous similar
works is that often the baryon component has been treated as a fixed shape
(rigid) potential. Here in general we do not adopt this approximation (since in
Paper I we wanted actually to study the expansion of leftover baryons), but we
evaluate its effect.

The organization of the paper is straightforward: in Section \ref{sec:method}
we describe the initial conditions and method for our simulations, whose result
are presented in Section \ref{sec:results}. The implications for our
understanding of ETGs formation are discussed in Section \ref{sec:discussion}.

%forse si potrebbero rivdiscutere un po' i risultati Ragone e Granato per i
%barioni nei profili con contrazione preliminare dm.

\begin{figure*}
\centerline{\includegraphics[width=14cm]{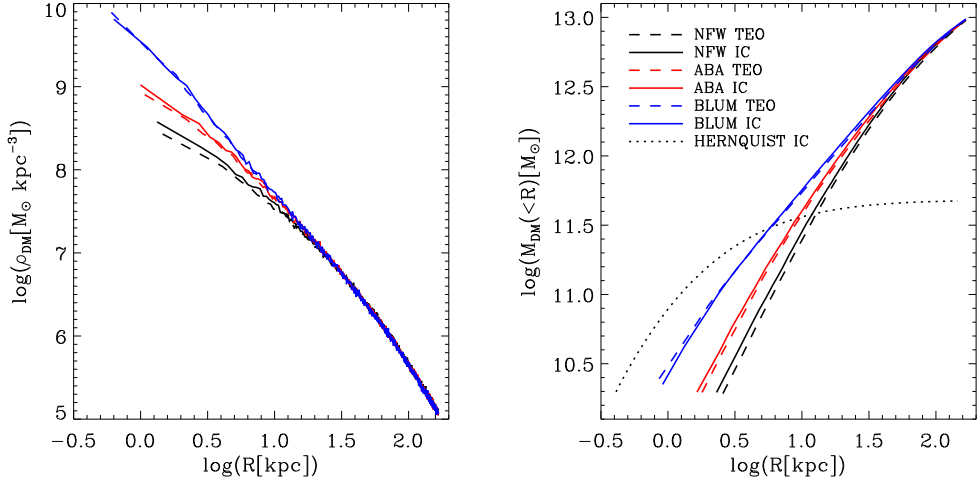}}
\vspace{-7cm}
  \caption{
The curve marked NFW TEO represents the typical density run of DM haloes
produced by gravity only simulations (Navarro, Frenk \& White 1997), whilst ABA
TEO and BLUM TEO are intended to keep into account the contraction resulting
from galaxy formation, as described in  Section \ref{sec:method}. After setting
these theoretical initial conditions (dashed lines in both panels) we let each
system to evolve, for 0.4 Gyr, to an equilibrium configuration.  These are
considered the initial conditions (IC) for the numerical experiments before
forcing baryon mass loss. The density and mass profiles are shown in the left
and right panel, respectively. In addition we show in the right panel the mass
Hernquist profile corresponding to the baryonic component. } \label{fig:ini}
\end{figure*}

\section{Numerical method and setup}
\label{sec:method}

The setup of the simulations used in this paper is similar to that described in
Paper I, to which we refer for a few more details. The main difference is that
here we run also simulations in which the initial conditions take into account
the possible halo contraction due to baryon condensation. Moreover, we
evaluated the effect of the approximation of treating the baryons as a fixed
shape potential.

The purpose of the simulations is to investigate the evolution of
collision-less particles (stars and DM) under a change of gravitational
potential due to a loss of baryonic mass of the system. In general, the
escaping mass can represent either the gas which has not been converted into
stars during the star forming phase of the spheroid (ejected by feedback driven
galactic winds), or the mass lost from stars in form of stellar winds and SNae
explosions (which is likely to escape the ETGs potential wells). In any case,
we assume as given, and due to  causes not included in our physical treatment
(such as SNae and AGN feedbacks, or stellar evolution), the temporal dependence
of this mass loss (Eq.~\ref{eq:massloss}), which we put by hand, and we
simulate the ensuing dynamical evolution of collision-less mass distributions.
Therefore we don't have to treat the gas dynamics.

We used the public version of the code \textsl{GADGET-2} (Springel et al.\
2005) to perform simulations with $10^6$ and $5\times 10^{6}$ particles, in the
gravity only mode. None of the presented results shows any noticeable
difference in the two cases, which assures us that the mass resolution is
sufficient for the purposes of the present study. Half of the particles are
used to sample the baryonic and dark matter components respectively, with a
softening of 0.007 and 0.35 kpc respectively. We checked that our results are
not affected by significant variations in these choices.

%Goodwin 06 tests show that the results are insensitive for any reasonable
%values.
%This convergence is unsurprising as the simulations follow the violent
%relaxation of the cluster to a new equilibrium, a situation in which
%two-body encounters are fairly unimportant and it is the bulk behaviour of
%the potential that dominates the evolution.

The density distribution of DM particles is initially (i.e. even before
computing the effects of baryonic contraction; see below) assumed to follow the
standard NFW (Navarro, Frenk \& White 1997) shape
\begin{equation}
\rho_{\rm DM}(r)={M_{\rm vir, DM}\over 4\pi\, R_{\rm
vir}^3}\,{c^2\,g(c)\over \hat r\, (1+c\, \hat r)^2}~,
\label{eq:nfw}
\end{equation}
where $M_{\rm vir, DM}$ is the halo virial mass in DM (the DM mass inside
$R_{\rm vir}$), $\hat r= r/R_{\rm vir}$, $c$ is the concentration parameter and
$g(c)\equiv [\ln(1+c)-c/(1+c)]^{-1}$.

%The virial radius, {\rev $R_{\rm vir}$, is by definition}  the radius within which
%the mean density is $\Delta_{\rm vir}(z)$ times the mean matter density of
%the universe $\rho_{u}(z)$. As such, it depends on $M_{\rm vir}$, on the
%virialization redshift of the halo and on the adopted cosmology. To
%compute it, we use the same, quite standard, relationships adopted by G04
%(their equations 1 and 2).

The virial radius $R_{\rm vir}$ is by definition the radius within which the
mean density is $\Delta_{\rm vir}(z_{vir})$ times the mean matter density of
the universe $\rho_{u}(z_{vir})$ at virialization redshift $z_{vir}$:
\begin{equation}
R_{\rm vir}=\left[\frac{3}{4 \pi} \frac{M_{\rm vir}}{\Delta_{\rm
vir}(z_{vir})\rho_{u}(z_{vir})}
\right]^{1/3}, \label{Rvir}
\end{equation}
The overdensity $\Delta_{\rm vir}(z)$, for a flat cosmology, can be
approximated by
\begin{equation}
\Delta_{\rm vir}(z) \simeq \frac{(18 \pi^2 +82x-39x^2)}{\Omega(z)},
\label{overdensity}
\end{equation}
where $x= \Omega(z)-1$ and $\Omega(z)$ is the ratio of the mean matter density
to the critical density at redshift $z$ (Bryan \& Norman 1998). The
corresponding mass distribution is written
\begin{equation}
M_{\rm DM}(<r)=M_{\rm vir,  DM}\, g(c)\,
\left[\ln{(1+c \, \hat r)}-{c \, \hat r\over
(1+c\, \hat r)}\right]~.
\label{eq:nfw_mass}
\end{equation}

For the collisionless baryonic particles (representing the potential of both
stars as well as gas before expulsion), we assume that, as a result of the
assembly of the central galaxy, they settle on an Hernquist (1990) profile,
which provides a reasonable description of stellar density in spheroids:
\begin{equation}
\rho_{\rm B}(r)={M_{\rm B}\over 2 \pi } {a \over r} {1 \over (r+a)^3}~;
\label{eq:her}
\end{equation}
where $M_{\rm B}$ is the total baryonic mass. The corresponding mass
distribution is
\begin{equation}
M_{\rm B}(<r)=M_{\rm B}\, \left({ r \over r+a}\right)^2~;
\label{eq:her_mass}
\end{equation}
so that the half-mass radius is related to the scale radius $a$ by
$R_{1/2}=(1+\sqrt{2})\, a$ and, assuming a mass to light ratio independent of
$r$, the effective radius is $R_{e}\simeq 1.81 a$.

As discussed in the introduction, the process of baryon condensation in the
centre of the DMHs is expected to contract to some extent the DM distribution.
Since, at variance with Paper I, the target of this study is precisely to
evaluate the effect of baryon expulsion on DMH density profiles, we performed
also runs including an estimate of the contraction in the initial conditions,
by adopting the Abadi et al.\ (2010) prescriptions. They found that a simple
formula captures the average behavior of their simulations:
\begin{equation}
r_f/r_i=1+\alpha\left[ (M_i / M_f)^n -1 \right]
\label{eq:aba}
\end{equation}
where $n$ and $\alpha$ are parameters (see below). This equation relates
the initial (i.e.\ before baryon condensation) and final radii of the
spheres containing the same amount of DM, to the {\it total} masses $M_f$
and $M_i$ within the same spheres. Since
\begin{equation}
\frac{M_f}{M_i}=\frac{M_{f,DM}(<r_f)+M_{f,bar}(<r_f)}{M_{i,DM}(<r_i)+M_{i,bar}(<r_i)}
\end{equation}
where $M_{f,DM}(<r_f)=M_{i,DM}(<r_i)$, by definition of $r_i$ and $r_f$. Once
it is assumed an initial density distribution for DM and baryons and a final
density distribution for baryons, Eq.~\ref{eq:aba} is an implicit equation for
$r_f$ given an $r_i$, that can be solved numerically to obtain the final mass
distribution of DM, $M_{f,DM}(<r)$. As stated by Abadi et al.\ (2010), their
results are well described setting $n=2$ and $\alpha=0.3$. Moreover the
contraction predicted by Blumenthal et al.\ (1986), as well as that found in
simulations by Gnedin et al.\ (2004), are well described setting $n=1$, and
$\alpha=1$ or $\alpha=0.73$ respectively. When exploiting Eq.~\ref{eq:aba}, we
describe the initial mass distribution of DM and the final mass distribution of
baryons with Eq.~\ref{eq:nfw_mass} and Eq.~\ref{eq:her_mass}. Also, we assume
that before contraction baryons follow the same density distribution of DM
(Eq.~\ref{eq:nfw}), rescaled by the cosmic baryon fraction we assume
$f_b=\Omega_{bar}/\Omega_{m}=0.17$ (consistent with CMB WMAP-7 constraints,
Komatsu et al.\ 2009).

In the following, unless otherwise specified, by {\it dynamical time} $t_{\rm
dyn}$ we mean the initial (i.e.\ before any mass loss and expansion) dynamical
time, computed at baryon $R_{1/2}$.
\begin{equation}
t_{\rm dyn}=\left[R_{1/2}^3\over 2 G \left(M_{\rm B}/2+M_{\rm
DM}(<R_{1/2})\right)
\right]^{1/2}
\label{eq:tdyn}
\end{equation}
Even though in this paper our focus is on the evolution of the DM component of
the system, we maintain the same definition of dynamical time as in Paper I,
since the effects under study are confined to the inner region of the DM
distribution.

\begin{figure*}
  \centerline{\includegraphics[width=16cm, height=14cm]{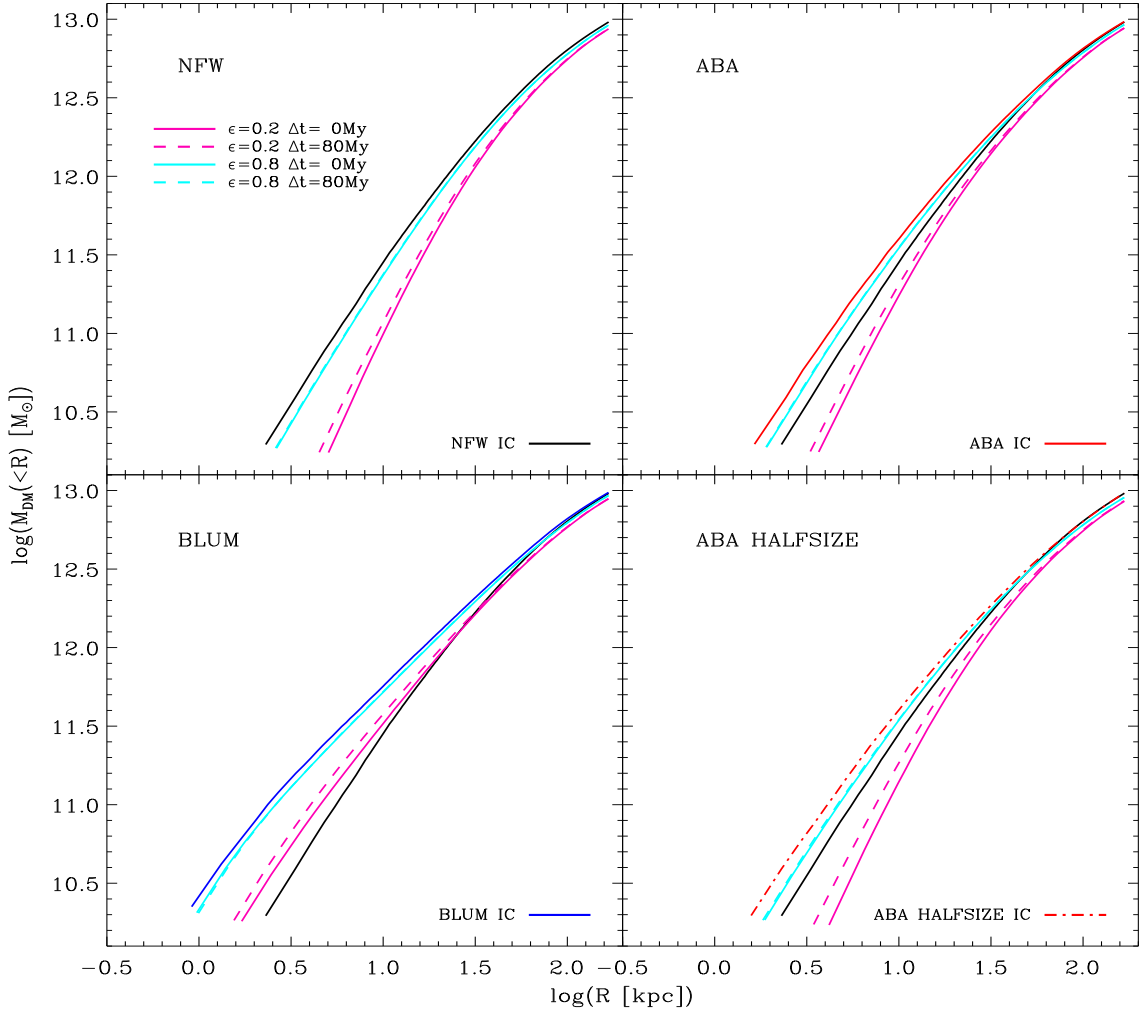}}
  \caption{
Comparison between the initial and final mass profiles. The latter is the new
equilibrium configuration reached $\sim 0.2$ Gyr after the baryonic mass loss
(see Fig.\ \ref{fig:evo}). Each panel refers to simulations performed adopting
a different initial density profile, as indicated by the inner label, which is
plotted with the same color as in previous figure. ABA HALFSIZE IC (bottom
right panel) is obtained as ABA IC, but halving the Hernquist scale radius of
the baryonic component. The final profiles for $\epsilon=0.2$ and
$\epsilon=0.8$ are in magenta and cyan respectively; solid lines are for
$\Delta t=0$ Myr (instantaneous baryon expulsion) and dashed lines for $\Delta
t=80$ Myr (barely distinguishable for the lower mass loss $\epsilon=0.8$). In
all panels we plot for reference also the standard NFW profile (black solid
line).} \label{fig:esp}
\end{figure*}

\begin{figure*}
  \centerline{\includegraphics[width=16cm, height=14cm]{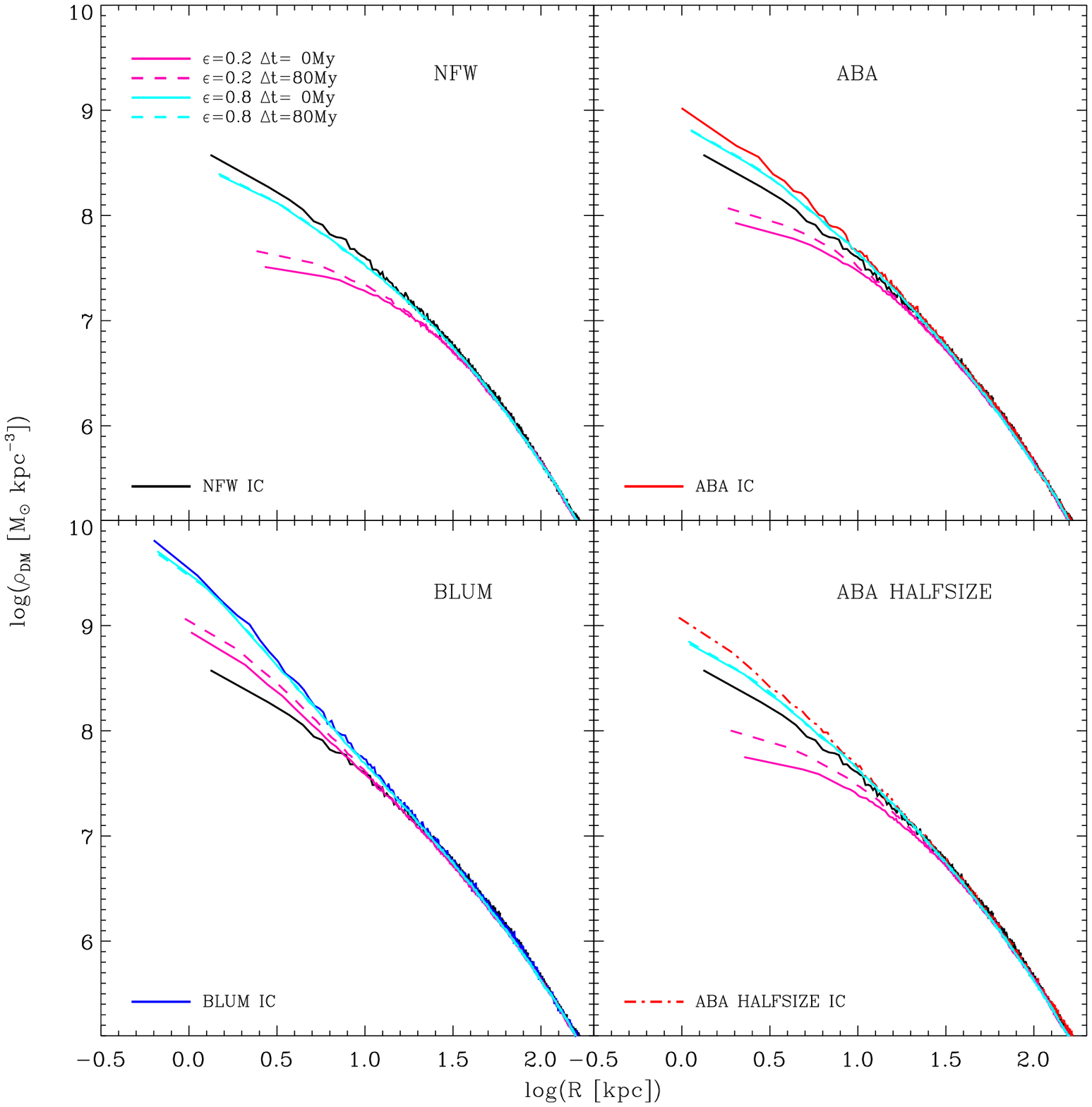}}
  \caption{The same as the previous figure but comparing the initial and final density profiles. The different panels
  correspond to different ICs as labeled in the plot.
 } \label{fig:esprho}
\end{figure*}

%For our standard initial conditions (see below) the contribution of DM to
%the mass inside $R_{1/2}$ amounts to $\simeq 20$\%. Thus $t_{\rm dyn}$ can
%be estimated (within 10\%) neglecting it:
%\begin{equation}
%t_{\rm dyn}\simeq 2.3 \left(R_e\over 1 {\rm kpc}\right)^{1.5} \left(M_{\rm
%B}\over
%10^{11}
%M_\odot\right)^{-0.5}
%\,\, {\rm Myr}
%\label{eq:tdynapp}
%\end{equation}

%Given the density runs, we obtain the 1D velocity dispersion by
%integrating the Jeans equation under the assumption of isotropic
%conditions:
%\begin{equation}
%\sigma_{X}^2(r)=-{1\over \rho_{X}(r)}\,\int_r^{\infty}{\rm
%d}{r'}\, {G\, M_{\rm TOT}(<r')\over
%r'^2}\, \rho_{X}(r')~,
%\end{equation}
%where $X$ stands for B or DM, and $M_{\rm TOT}(<r)=M_{\rm DM}(<r)+M_{\rm
%B}(<r)$. By evolving the particle system for several dynamical times, we
%get confident that it is actually in (quasi-)static statistical
%equilibrium.

Given the density profiles, we obtain the 1D velocity dispersion by integrating
the Jeans equation under the assumption of isotropic conditions. To generate
initial conditions, we randomly populate the system with baryonic and DM
particles, according to the density distributions Eq. \ref{eq:her} and Eq.\
\ref{eq:nfw}.  The particles velocities are randomly generated assuming local
Maxwellian distribution, with 1D velocity dispersion obtained by the solution
of Jeans equation.  Then, by numerically evolving the system for several
dynamical times, we obtain a (quasi-)static statistical equilibrium. This
latter step does not produce significant variations in the density profiles
(see discussion of Fig.\ \ref{fig:ini} in Section \ref{sec:results} for more
details).

Starting from this initial setup, we introduce a mass loss, by removing
exponentially during an ejection time interval $\Delta t$ a fraction
$1-\epsilon$ of the baryonic mass :
\begin{equation}
M_B(t)=M_{\rm B(t=0)}\,\exp\left({\ln \epsilon \over \Delta t} \, t\right)~
\label{eq:massloss}
\end{equation}
%For instance, this simple functional form provides an acceptable
%description of the gas removal due to QSO feedback in the G04
%semi-analytic model, with an ejection timescale of the order of 20-30 Myr
%for a wide range of the model parameters.

The mass loss is practically attained by decreasing correspondingly in time the
mass of the baryonic particles sampling the density field. After the end of the
mass loss period $\Delta t$, we let the system to evolve till it reaches a new
equilibrium configuration.

We wanted also to study the effect of a common approximation done when
evaluating the evolution of the DM distribution under a baryon mass loss,
namely to treat the baryonic component as a potential that changes in intensity
without any variation in shape (a rigid potential; e.g.\ Navarro et al.\ 1996;
Ogiya \& Mori 2011). For this purpose, we run also simulations in which the
positions of baryonic particles were not updated in time.

The reference value for the initial (i.e.\ before any mass loss) ratio of
virial mass (total mass within the virial radius) to baryonic mass is $M_{\rm
vir}/M_{\rm B(t=0)}=20$. {\rev This value is in keeping with the fraction
20\%-40\% of cosmic baryons that can cool and condense in the central region of
a large galactic DM halo at $z\gtrsim 1.5$, according to the prescriptions
adopted by semi analytic models and simulations of galaxy formation (e.g.\
Benson et al.\ 2001; Helly et al.\ 2003;  Granato et al.\ 2004; Cattaneo et
al.\ 2007; Viola et al.\ 2008). Moreover, assuming that after the initial
condensation the halo looses, due to feedback induced galactic winds, a
fraction between 20\% to 80\% of this "galactic" baryonic mass, it is left with
a baryon to DM content broadly consistent with estimates in the local universe
for large galactic haloes (Monster et al.\ 2010).}

We set $M_{\rm vir}=10^{13}{\rm M}_\odot$ in all simulations. Nevertheless, our
results apply to different values of $M_{\rm vir}$, provided that the ratios of
scale radii and masses in the two components (DM and baryons) are not changed,
and that the time is measured in units of dynamical time $t_{\rm dyn}
\propto\rho^{-1/2}$.

We adopt  a concentration parameter $c=4$, a typical value at galactic halo
formation (see Zhao et al. 2003; Klypin, Trujillo-Gomez, \& Primack 2010), and
$R_{\rm vir}\simeq 170$ kpc (from Eq.~\ref{Rvir} and Eq.~\ref{overdensity},
with $M_{\rm vir}=10^{13} {\rm M}_\odot$ and $z_{vir}=3$).

Finally, we set $a=1.5$ kpc ($R_e\simeq 2.7$ kpc). We refer the reader to
Section 3 of Paper I for motivations for this values. The initial (i.e.\ before
mass loss and expansion, Eq.\ \ref{eq:tdyn}) dynamical time as defined above is
$t_{\rm dyn}\approx 5$ Myr.

In summary, the parameters affecting the results of our simulations are the
ratio of mass between the total and baryonic components $M_{\rm vir}/M_{\rm
B(t=0)}$; the corresponding ratio of scale-lengths $R_{\rm vir}/a$; the
fraction of baryonic mass lost $(1-\epsilon)$, and the time $\Delta t$ over
which the loss occurs. We performed simulations covering broad ranges of the
latter two quantities, while in most runs we kept the former two at the
fiducial values reported above. We checked however that none of our qualitative
conclusion is affected by factor $\sim 2$ variations of them (Paper I and some
more discussion in Section \ref{sec:results}).

\begin{figure*}
\centerline{\includegraphics[width=14cm]{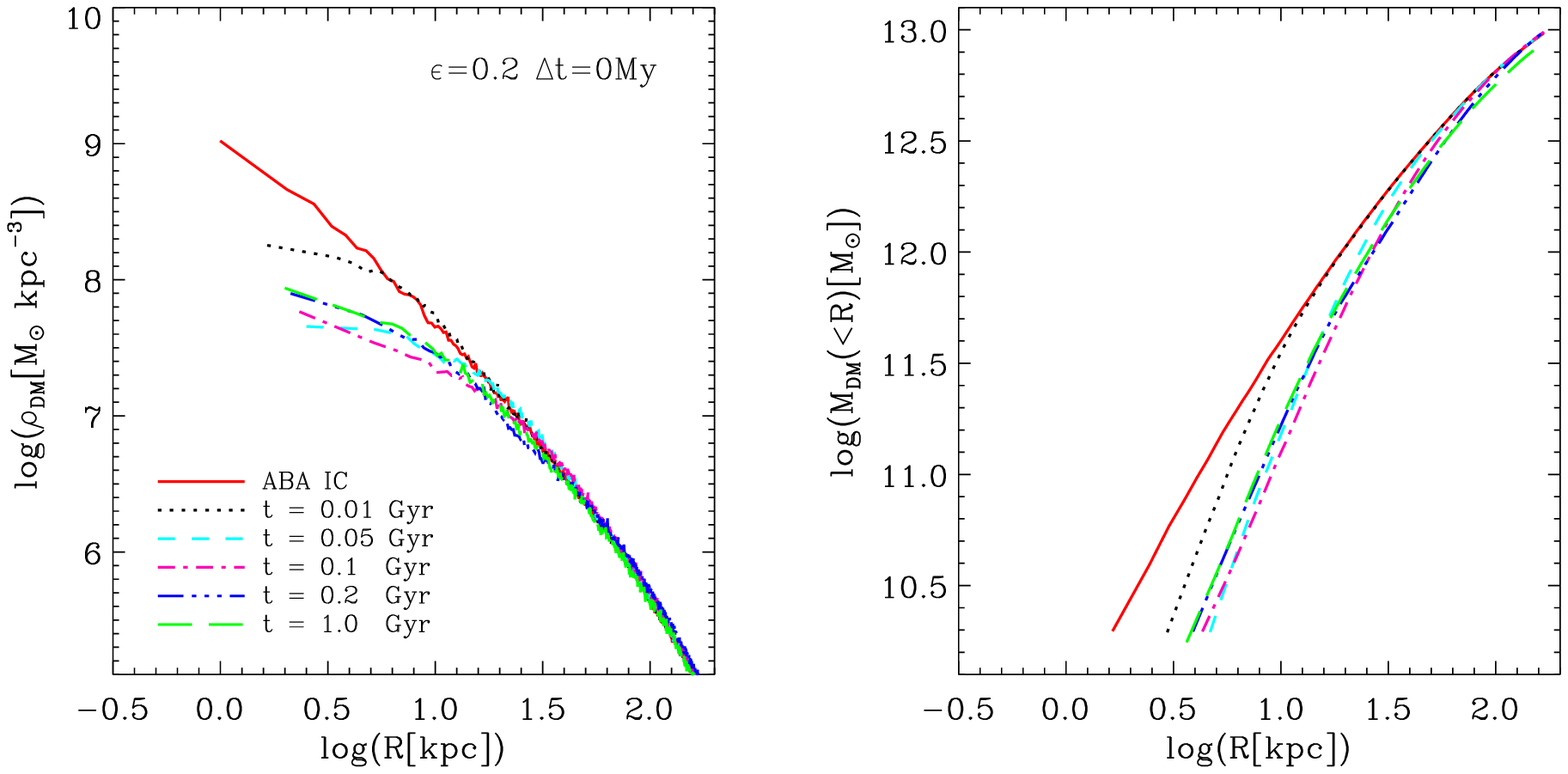}} \vspace{-7cm}
\caption{Time evolution, after the baryon mass loss, of the density (left
panel) and mass (right panel) DM profiles for one of our simulations, namely
the case ABA with  instantaneous mass loss $\Delta t=0$ Gyr and $\epsilon=0.2$.
The new  equilibrium is reached within $\sim 0.2$ Gyr, i.e. within 40 dynamical
times of the  central region, Eq. \ref{eq:tdyn}. This holds true for all the
other cases.  }
  \label{fig:evo}
\end{figure*}

\begin{figure*}
\centerline{\includegraphics[width=14cm]{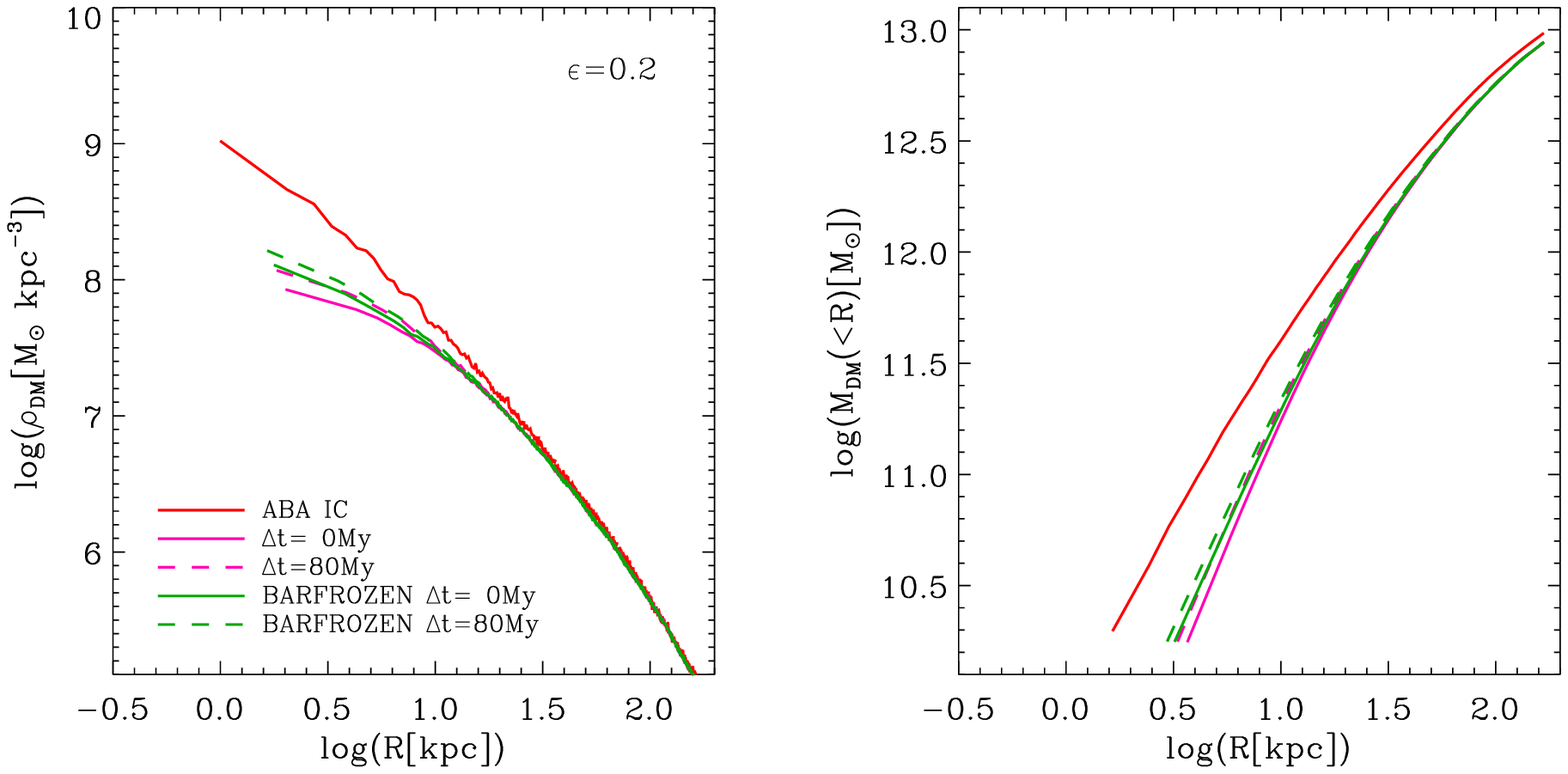}}
\vspace{-7cm}

\caption{Comparison between the new equilibrium DM density (left panel) and
mass (right panel) profiles after baryon mass loss, for $\epsilon=0.2$,
obtained letting the baryon particles to adjust their position according to the
change of potential (magenta lines), or keeping  their position fixed (green
lines). The former is the treatment adopted in this work, while the latter
corresponds to the common approximation of treating the baryons as a rigid
potential. } \label{fig:clava}
\end{figure*}
\section{Results}
\label{sec:results}

Fig.\ \ref{fig:ini} shows the initial equilibrium density and mass profiles for
the DM component, i.e.\ those adopted before forcing any loss of baryonic mass
(solid lines). The right panel also displays the adopted standard Hernquist
mass profile for the baryons. Here and in the following, the lower limit
of the density and mass profile plots equals the adopted gravitational
softening for the DM component, the density and the accumulated mass are
evaluated using radial bins containing $\simeq 500$ DM particles. These
initial profiles are obtained letting to evolve for several tens of dynamical
times the composite system, where for the DM component we adopted the pure NFW
mass profile (black dashed line), or the profiles implicitly given by
Eq.~\ref{eq:aba}, with values of the $n$ and $\alpha$ parameters adequate to
reproduce the contraction found by Abadi et al.\ (2010) in simulations (red
dashed line) or that analytically predicted by Blumenthal et al.\ (1986) (blue
dashed line), as detailed in Section \ref{sec:method}. We refer to these three
cases as NFW, ABA and BLUM respectively. The first and the last should be
regarded as extreme cases where the effect of contraction is negligible and
maximal, respectively, while the intermediate case, ABA, is likely to be more
realistic. The numerical evolution does not affect much the initial
configuration of DM. For instance, the local power law index of the density
distribution increases in the inner $\sim 10$ kpc by less than $0.1$, making it
slightly steeper. The effect decreases with increasing contraction of the DM
profiles, i.e.\ from NFW to BLUM. Kazantzidis, Magorrian \& Moore (2004)
suggested that the local Maxwellian assumption employed here (and in many other
works) to set up the initial conditions, can lead to spurious results in the
study of long term evolution of DM haloes. In particular they found that the
central cusp is quickly significantly reduced. In our case we find a much less
important, and opposite, trend. We attribute this to the presence of the
dominating baryonic component in the center of the halo, not included in the
study by Kazantzidis et al. Indeed, on one hand we run a pure DM test case,
which confirmed their result on the reduction of the central cusp. On the other
hand, we note that also Read \& Gilmore (2005), who considered composite baryon
plus DM systems as we did, found excellent agreement between initial conditions
generated with the Maxwellian approximation and with the alternative procedure
proposed by Kazantzidis et al.\ (2004).

In addition to the three cases NFW, ABA and BLUM, and in order to study the
dependence of our results on the parameters of the initial baryonic
configuration, we considered ICs in which we halved the scale radius of the
baryonic component (henceforth ABA HALFSIZE) or we doubled its mass. We then
correspondingly contract the DM halo as we did for ABA. The direction of these
variations has been elected in order to increase the effect of mass loss with
respect to the standard case. These two initial conditions are not shown in
Fig.\ \ref{fig:ini}, since they are almost indistinguishable from ABA IC.

Fig.\ \ref{fig:esp} and Fig.\ref{fig:esprho} display the effect of baryon mass
loss on the  initial mass distributions described above. The four panels in each figure
correspond to the ICs dubbed NFW, ABA, BLUM and ABA HALFSIZE. The new
equilibrium configurations, shown in the figures, are reached typically a few
tens of dynamical times (Eq. \ref{eq:tdyn}) after the end of the mass loss
period ($\Delta t$), as can be appreciated in Fig.\ \ref{fig:evo}. Whenever the fraction of
baryon lost is important (say $\gtrsim 50 \%$), the final DM profile is
significantly flatter in the center than the initial one. If this is the case,
the ensuing expansion more than counteracts the opposite contraction caused by
baryon condensation, at least that found in recent numerical simulations (Abadi
et al.\ 2010), see right panels. The effect is somewhat weaker when the DM
profile is initially more concentrated, as expected since in this case the
contribution of DM to the gravitational field in the center becomes more
important. For the same reason, the flattening effect in ABA HALFSIZE is
slightly enhanced with respect to ABA. Indeed, in the former ICs, the
contribution to the initial equilibrium of the baryons that we let to escape
during mass loss was higher. To a lesser extent, the same happens doubling the
initial baryon mass, but the difference is barely visible, so that we don't
plot this case.

The dependence on the timescale of the mass loss is weak. This finding is at
variance with respect to the result by Ogiya \& Mori (2011), who found that if
the loss is slow enough, the DM profiles are much less affected, up to the
point that after a while it returns very close to the initial conditions. We
have checked that this result is not due to the approximation used by Ogiya \&
Mori of treating the baryons as a fixed form potential (see below). This
difference could arise in part, but not entirely, from the more compact baryon
distribution adopted by these authors. Actually, we find a somewhat more
evident dependence on timescale when the baryon component is initially more
concentrated (ABA HALFSIZE; lower right panels). It is also interesting to
note that the expansion of the baryonic mass distribution is much more affected
than that of the DM by the timescale of expulsion  (see Paper I for details).

In Fig. \ref{fig:evo} we show the time evolution of the density (left panel)
and the mass (right panel) profiles of a representative model, after the baryon
mass loss. The chosen run corresponds to the case ABA with  instantaneous mass
loss $\Delta t=0$ Gyr and $\epsilon=0.2$. The figure shows that the profile at
remains stable after $\sim 0.2$ Gyr. The new  equilibrium is reached within 40
dynamical times of the central region (Eq. \ref{eq:tdyn}). This holds true for
all the other simulated cases.

A commonly made approximation in evaluating the effect of baryon expulsion (or
condensation) on the DM halo is to treat the baryonic component as a potential
that changes in intensity without any variation in shape. Fig.\ \ref{fig:clava}
illustrates that this approximation leads to some underestimate of the slope
flattening and of its (weak) dependence on the expulsion timescale. This is due
to the fact that the mass loss actually causes a significant broadening of the
baryon concentration in the centre, which changes the shape of the potential
and is more important for shorter timescales (Paper I). Therefore, in the full
computation, the DM distribution is flattened not only by the decrease of
gravitational force of baryons, but also by the (timescale dependent) outside
dragging, due to the expansion of the leftover baryonic matter.

\section{Discussion and conclusions}
\label{sec:discussion}

A well know general prediction of cosmological, gravity only, simulations is
that DM haloes should have cuspy density profiles, essentially independently of
the mass scale\footnote{note that this feature has been found recently also in
self-similar analytic models for the halo collapse (e.g., Lapi \& Cavaliere
2011).}. Observations at small to medium galactic scales (dwarf galaxies, disc
dominated LSB galaxies, as well as normal spirals) have demonstrated that this
is not the case. At cluster scales, where DM is gravitationally subdominant in
the central region, the situation is instead far from clear, with several
claims for cored (e.g.\ Ricthler et al.\ 2011) as well as for cuspy (e.g.\
Zitrin et al.\ 2011) density profiles. The former mismatch is widely ascribed
to the back-reaction of baryons, whose evolution to form galaxies is driven by
various non-gravitational processes, on DM particles. The intermediate regime
of large ETGs, the subject of our work, has received so far much less
attention, both from the observational as well as the theoretical point of
view. As for the former, again there are interpretations favoring a cored
(e.g.\ Memola et al.\ 2011) as well as a cuspy (Tortora et al.\ 2010,
Sonnenfeld et al.\ 2011) density profile. Firm conclusions are however severely
plagued by the degeneracy between the Initial Mass Function (IMF) and the DM
density profiles (Treu et al.\ 2010).

In the present paper, we have elucidated that an important gas removal during
the early evolution of ETGs, should leave as a byproduct sizeable signatures
also on the inner profile of their DM haloes. A similar ejection is required by
most galaxy formation models aiming to explain the basic properties of these
systems, such as their chemical properties, low baryon content or luminosity
function (e.g.\ Benson et al.\ 2003; Granato et al.\ 2004; Pipino, Silk \&
Matteucci 2009; Duffy et al.\ 2010), and it is commonly, but not always,
ascribed to {\it QSO mode} AGN feed-back. The DM density profile ends up to be
significantly less concentrated than NFW, unless the prior (opposite)
contraction generated by baryon collapse and condensation has been very
efficient, and probably unrealistic, i.e.\ closer to that estimated on the
basis of approximate analytical treatments (e.g.\ Blumenthal et al.\ 1986),
than to that found in most cosmological simulations (e.g.\ Abadi et al.\ 2010;
Gnedin et al.\ 2011).

Moreover, it has been pointed out that stellar feedback can weaken the DM cusp
in dwarf and spiral galaxies not only by means of gas removal from the galaxy
potential well, but also as a consequence of oscillations of the potential
generated by bulk motions of gas within the galaxy (Mashchenko et al.\ 2006,
2008; {\rev Potzen \& Governato 2012}; Macci{\`o} et al.\ 2012). As for the AGN
feedback, similar fluctuations could be induced, {\rev for instance}, by non
isotropic gas removal. Although, for simplicity and lack of theoretical
understanding, models of galaxy formation treat AGN feedback by means of
isotropic sub-grid prescriptions, the case for preferential directions for the
effect of AGN activity on its environment is actually strong. Indeed, all our
knowledge of the AGN phenomenon points to a non isotropic structure, including
accretion discs, jets, ionizations cones and dust tori. More specifically,
recent observational evidence directly indicates non isotropic quasar driven
gas removal (e.g.\ Cano-Diaz et al.\ 2011). To explore properly this effect
would require more complex numerical experiments, possibly including a
treatment of gas dynamics. The effort seems somewhat premature at present,
given the lack of understanding of how AGN gas removal works, and we leave it
for future investigations\footnote{Peirani et al.\ (2008) performed idealized
numerical experiments partly in this spirit, in which periodic AGN activity is
assumed to cause bulk oscillations, rather than removal, of gas. {\rev See also
discussion in Potzen \& Governato 2012.}}. In any case, it is worth noticing
that this possible additional process could even enhance the flattening effect
of AGN feedback on the DM distribution estimated here.

{\rev Martizzi et al.\ (2011) suggested that several mechanisms contribute to
the formation of the $\sim 10$ kpc core in the Brightest Cluster Galaxy of a
simulated Virgo-like galaxy cluster ($M_{vir} \simeq 10^{14} M_\odot$), found
when (and only when) sub-grid models for the growth of SMBHs and the ensuing
feedback are included in their hydro-simulation. Our idealized numerical
experiments elucidates and quantifies the important contribution of one of
these mechanismsa, namely the ejection of baryonic matter (representing gas).}

In conclusion, cuspy density profiles in ETGs, tentatively inferred from some
recent observation, could be difficult to reconcile with an effective AGN (or
stellar) feedback, in particular that believed to cause massive galactic winds
during the early evolution of these systems, making them {\it red and
dead}.

%Read \& Gilmore sempre perdono 95\% e nella stessa scala temporale, 1
%baryon crossing time (impulsivo). tolgono i barioni con altra tecnica,
%dando una velocita' random. Prima fanno contrazione numerica o analitica,
%la prima aumentando la massa su tempi scala pero' sempre un po' rapidini
%(1-15-40 tcross). il set e' B1 B2 B3 B4. In tutti i casi i barioni nella
%galassia sono inchiodati e' cosi?.

\section*{Acknowledgments}
C.R-F.\ and G.L.G.\ acknowledge warm hospitality by INAF-Trieste and
IATE-C\'ordoba, respectively, during the development of the present work. We
thank A.\ Lapi, P.\ Salucci, J.\ Navarro
and A. Meza for several useful
discussions on the topic of this work. This work has been partially supported
by the Consejo de Investigaciones Cient\'{\i}ficas y T\'ecnicas de la
Rep\'ublica Argentina (CONICET), by the Secretar\'{\i}a de Ciencia y T\'ecnica
de la Universidad Nacional de C\'ordoba (SeCyT) and by the European Commission's
Framework Programme 7, through the International Research Staff Exchange Scheme
LACEGAL.

{}

\clearpage


\begin{thebibliography}{}

\bibitem[\protect\citeauthoryear{Abadi et al.}{2010}]{2010MNRAS.407..435A}
Abadi M.~G., Navarro J.~F., Fardal M., Babul A., Steinmetz M., 2010, MNRAS,
407, 435

%\bibitem[]{}Baumgardt, H., \& Kroupa, P. 2007, MNRAS, 380, 1589

\bibitem[\protect\citeauthoryear{Benson et
    al.}{2003}]{2003ApJ...599...38B} Benson A.~J., Bower R.~G., Frenk
    C.~S., Lacey C.~G., Baugh C.~M., Cole S., 2003, ApJ, 599, 38

\bibitem[\protect\citeauthoryear{Benson et al.}{2001}]{2001MNRAS.320..261B}
    Benson A.~J., Pearce F.~R., Frenk C.~S., Baugh C.~M., Jenkins A., 2001,
    MNRAS, 320, 261

%\bibitem[\protect\citeauthoryear{Benson}{2010}]{2010PhR...495...33B}
%    Benson A.~J., 2010, PhR, 495, 33

%\bibitem[]{}Bezanson, R., et al. 2009, ApJ, 697, 1290

%\bibitem[]{}Biermann, P., \& Shapiro, S.L. 1979, ApJ, 230, L33

\bibitem[\protect\citeauthoryear{Blumenthal et
al.}{1986}]{1986ApJ...301...27B} Blumenthal G.~R., Faber S.~M., Flores R.,
Primack J.~R., 1986, ApJ, 301, 27

%\bibitem[]{}Boily, C.M., \& Kroupa, P. 2003, MNRAS, 338, 673

%\bibitem[]{}Boylan-Kolchin, M., Ma, C.-P., \& Quataert, E. 2006, MNRAS,
%    369, 1081

%\bibitem[]{}Boylan-Kolchin, M., Ma, C.-P., \& Quataert, E. 2008, MNRAS,
%    383, 93

\bibitem[\protect\citeauthoryear{Brammer et
al.}{2011}]{2011ApJ...739...24B} Brammer G.~B., et al., 2011, ApJ, 739, 24

\bibitem[\protect\citeauthoryear{Bryan
& Norman}{1998}]{1998ApJ...495...80B} Bryan G.~L., Norman M.~L., 1998,
 ApJ, 495, 80

%\bibitem[]{}Buitrago, F., et al. 2008, ApJ, 687, L61

\bibitem[\protect\citeauthoryear{Buote
\& Humphrey}{2012}]{2012ASSL..378..235B} Buote D.~A., Humphrey P.~J., 2012, ASSL, 378, 235

%\bibitem[\protect\citeauthoryear{Cassata et
%    al.}{2010}]{2010ApJ...714L..79C} Cassata P., et al., 2010, ApJ, 714,
%    L79

\bibitem[\protect\citeauthoryear{Cano-Diaz et
al.}{2011}]{2011arXiv1112.3071C} Cano-Diaz M., Maiolino R., Marconi A.,
Netzer H., Shemmer O., Cresci G., 2011, arXiv, arXiv:1112.3071

\bibitem[\protect\citeauthoryear{Cattaneo et al.}{2007}]{2007MNRAS.377...63C}
    Cattaneo A., et al., 2007, MNRAS, 377, 63

\bibitem{}Cattaneo, A., et al. 2006, MNRAS, 370, 1651

%\bibitem[]{}Cimatti, A., et al. 2008, A\&Ap, 482, 21

\bibitem[\protect\citeauthoryear{Ciotti, Ostriker, \&
    Proga}{2009}]{2009ApJ...699...89C} Ciotti L., Ostriker J.~P., Proga
   D., 2009, ApJ, 699, 89

%\bibitem[\protect\citeauthoryear{Ciotti \& van
%    Albada}{2001}]{2001ApJ...552L..13C} Ciotti L., van Albada T.~S., 2001,
%    ApJ, 552, L13

%\bibitem[]{}Cook, M., Lapi, A., \& Granato, G.L. 2009, MNRAS, 397, 534

%\bibitem[]{}Coppin, K., et al. 2006, MNRAS, 372, 1621

%\bibitem[]{}Cowie, L.L., Songaila, A., Hu, E.M., \& Cohen, J.G. 1996, AJ,
%    112, 839

%\bibitem[]{}Cresci, G.,  et al. 2009, ApJ, 697, 115

%\bibitem[]{}Daddi, E., et al. 2009, ApJ, 695, L176

%\bibitem[\protect\citeauthoryear{Daddi et al.}{2005}]{2005ApJ...626..680D}
%    Daddi E., et al., 2005, ApJ, 626, 680

%\bibitem[]{}Damjanov, I., et al. 2009, ApJ, 695, 101

\bibitem[\protect\citeauthoryear{de Blok}{2010}]{2010AdAst2010E...5D} de
Blok W.~J.~G., 2010, AdAst, 2010

\bibitem[\protect\citeauthoryear{de Souza et
al.}{2011}]{2011MNRAS.415.2969D} de Souza R.~S., Rodrigues L.~F.~S., Ishida
E.~E.~O., Opher R., 2011, MNRAS, 415, 2969

%\bibitem[]{}Drory, N., et al. 2005, ApJ, 619, L131

\bibitem[\protect\citeauthoryear{Dubinski
\& Carlberg}{1991}]{1991ApJ...378..496D} Dubinski J., Carlberg R.~G., 1991, ApJ, 378, 496

\bibitem[\protect\citeauthoryear{Duffy et al.}{2010}]{2010MNRAS.405.2161D}
Duffy A.~R., Schaye J., Kay S.~T., Dalla Vecchia C., Battye R.~A., Booth
C.~M., 2010, MNRAS, 405, 2161

%\bibitem[]{}Dye, S., et al. 2008, MNRAS, 386, 1107

%\bibitem[]{}Fan, L., Lapi, A., De Zotti, G., \& Danese, L. 2008, ApJ, 689,
%    L101

%\bibitem[]{}Fan, L., Lapi, A., Bernardi, M., Bressan, A., De Zotti, G., \&
%    Danese, L. 2010, ApJ, 718, 1460

\bibitem[\protect\citeauthoryear{Fabian}{1999}]{1999MNRAS.308L..39F} Fabian
A.~C., 1999, MNRAS, 308, L39

\bibitem[\protect\citeauthoryear{Fontanot et
al.}{2009}]{2009MNRAS.397.1776F} Fontanot F., De Lucia G., Monaco P.,
Somerville R.~S., Santini P., 2009, MNRAS, 397, 1776

\bibitem[\protect\citeauthoryear{Gnedin et al.}{2011}]{2011arXiv1108.5736G}
Gnedin O.~Y., Ceverino D., Gnedin N.~Y., Klypin A.~A., Kravtsov A.~V.,
Levine R., Nagai D., Yepes G., 2011, arXiv, arXiv:1108.5736

\bibitem[\protect\citeauthoryear{Gnedin et al.}{2004}]{2004ApJ...616...16G}
Gnedin O.~Y., Kravtsov A.~V., Klypin A.~A., Nagai D., 2004, ApJ, 616, 16

\bibitem[\protect\citeauthoryear{Gnedin
\& Zhao}{2002}]{2002MNRAS.333..299G} Gnedin O.~Y., Zhao H., 2002, MNRAS, 333, 299

\bibitem[\protect\citeauthoryear{Governato et
al.}{2010}]{2010Natur.463..203G} Governato F., et al., 2010, Natur, 463,
203

\bibitem[\protect\citeauthoryear{Granato et
al.}{2004}]{2004ApJ...600..580G} Granato G.~L., De Zotti G., Silva L.,
Bressan A., Danese L., 2004, ApJ, 600, 580

\bibitem[\protect\citeauthoryear{Granato et
al.}{2001}]{2001MNRAS.324..757G} Granato G.~L., Silva L., Monaco P.,
Panuzzo P., Salucci P., De Zotti G., Danese L., 2001, MNRAS, 324, 757

\bibitem[\protect\citeauthoryear{Helly et al.}{2003}]{2003MNRAS.338..913H}
    Helly J.~C., Cole S., Frenk C.~S., Baugh C.~M., Benson A., Lacey C., Pearce
    F.~R., 2003, MNRAS, 338, 913

\bibitem[\protect\citeauthoryear{Hernquist}{1990}]{1990ApJ...356..359H}
Hernquist L., 1990, ApJ, 356, 359

\bibitem[\protect\citeauthoryear{Inoue
\& Saitoh}{2011}]{2011MNRAS.418.2527I} Inoue S., Saitoh T.~R., 2011, MNRAS, 418, 2527

\bibitem[]{}Johansson, P.H., Naab, T., \& Burkert, A. 2009, ApJ, 690, 802

%\bibitem[]{}Joung, M.R., Cen, R., \& Bryan, G.L. 2009, ApJ, 692, L1

%\bibitem[]{}J{\o}rgensen, I., Franx, M., \& Kjaergaard, P. 1993, ApJ, 411,
%    34

\bibitem[\protect\citeauthoryear{Kazantzidis, Magorrian,
\& Moore}{2004}]{2004ApJ...601...37K} Kazantzidis S., Magorrian J., Moore B., 2004, ApJ, 601, 37

%\bibitem[\protect\citeauthoryear{Khochfar \&
%    Silk}{2006}]{2006ApJ...648L..21K} Khochfar S., Silk J., 2006, ApJ,
%    648, L21

\bibitem[\protect\citeauthoryear{Klypin, Trujillo-Gomez, \&
    Primack}{2010}]{2010arXiv1002.3660K} Klypin A., Trujillo-Gomez S.,
    Primack J., 2010, arXiv, arXiv:1002.3660

\bibitem[\protect\citeauthoryear{Komatsu et
al.}{2009}]{2009ApJS..180..330K} Komatsu E., et al., 2009, ApJS, 180, 330

%\bibitem[\protect\citeauthoryear{La Barbera et
%    al.}{2009}]{2009AJ....137.3942L} La Barbera F., de Carvalho R.~R., de
%    la Rosa I.~G., Sorrentino G., Gal R.~R., Kohl-Moreira J.~L., 2009, AJ,
%    137, 3942

%\bibitem[]{}Lapi, A., et al. 2006, ApJ, 650, 42

%\bibitem[]{}L{\'{\i}}pari, S., et al. 2009, MNRAS, 398, 658

%\bibitem[]{}Longhetti, M., et al. 2007, MNRAS, 374, 614

%\bibitem[]{}Maier, C., et al. 2009, ApJ, 694, 1099

%\bibitem[]{}Maller, A.H., et al. 2006, ApJ, 647, 763

%\bibitem[]{}Mancini, C., et al. 2010, MNRAS, 401, 933

%\bibitem[\protect\citeauthoryear{McGrath et
%    al.}{2008}]{2008ApJ...682..303M} McGrath E.~J., Stockton A., Canalizo
%    G., Iye M., Maihara T., 2008, ApJ, 682, 303

\bibitem[\protect\citeauthoryear{Lapi
\& Cavaliere}{2011}]{2011ApJ...743..127L} Lapi A., Cavaliere A., 2011, ApJ, 743, 127

\bibitem[\protect\citeauthoryear{Macci{\`o} et
al.}{2012}]{2012ApJ...744L...9M} Macci{\`o} A.~V., Stinson G., Brook C.~B.,
Wadsley J., Couchman H.~M.~P., Shen S., Gibson B.~K., Quinn T., 2012, ApJ,
744, L9

\bibitem[\protect\citeauthoryear{Martizzi et al.}{2011}]{2011arXiv1112.2752M}
    Martizzi D., Teyssier R., Moore B., Wentz T., 2011, arXiv, arXiv:1112.2752

\bibitem[\protect\citeauthoryear{Mashchenko, Wadsley,
\& Couchman}{2008}]{2008Sci...319..174M} Mashchenko S., Wadsley J., Couchman H.~M.~P., 2008, Sci, 319, 174
\bibitem[\protect\citeauthoryear{Mashchenko, Couchman,
\& Wadsley}{2006}]{2006Natur.442..539M} Mashchenko S., Couchman H.~M.~P., Wadsley J., 2006, Natur, 442, 539

\bibitem[\protect\citeauthoryear{Memola, Salucci,
\& Babi{\'c}}{2011}]{2011A&A...534A..50M} Memola E., Salucci P., Babi{\'c} A., 2011, A\&A, 534, A50

\bibitem{}Monaco, P., Fontanot, F., \& Taffoni, G. 2007, MNRAS, 375, 1189

\bibitem[]{}Moster, B.P., et al. 2010, 710, 903

%\bibitem[]{}Naab, T., Johansson, P.H., Ostriker, J.P., \& Efstathiou, G.
%    2007, ApJ, 658, 710

%\bibitem[]{}Naab, T., Johansson, P.H., \& Ostriker, J.P. 2009, ApJ, 699,
%    L178

\bibitem[]{}Navarro, J.F., Frenk, C.S., \& White, S.D.M. 1997, ApJ, 490, 493

\bibitem[\protect\citeauthoryear{Navarro, Eke,
\& Frenk}{1996}]{1996MNRAS.283L..72N} Navarro J.~F., Eke V.~R., Frenk C.~S., 1996, MNRAS, 283, L72

\bibitem[\protect\citeauthoryear{Newman et al.}{2011}]{2011arXiv1110.1637N}
Newman A.~B., Ellis R.~S., Bundy K., Treu T., 2011, arXiv, arXiv:1110.1637

%\bibitem[\protect\citeauthoryear{Nipoti, Londrillo, \&
%    Ciotti}{2003}]{2003MNRAS.342..501N} Nipoti C., Londrillo P., Ciotti
%    L., 2003, MNRAS, 342, 501

\bibitem[\protect\citeauthoryear{Ogiya
\& Mori}{2011}]{2011ApJ...736L...2O} Ogiya G., Mori M., 2011, ApJ, 736, L2

%\bibitem[\protect\citeauthoryear{Onodera et
%    al.}{2010}]{2010ApJ...715L...6O} Onodera M., et al., 2010, ApJ, 715,
%    L6

%\bibitem[\protect\citeauthoryear{Oser et al.}{2010}]{2010ApJ...725.2312O}
%    Oser L., Ostriker J.~P., Naab T., Johansson P.~H., Burkert A., 2010,
%    ApJ, 725, 2312

\bibitem[\protect\citeauthoryear{Pasetto et
al.}{2010}]{2010A&A...514A..47P} Pasetto S., Grebel E.~K., Berczik P., Spurzem R., Dehnen W., 2010, A\&A, 514, A47

\bibitem[\protect\citeauthoryear{Peirani, Kay,
\& Silk}{2008}]{2008A&A...479..123P} Peirani S., Kay S., Silk J., 2008, A\&A, 479, 123

\bibitem[\protect\citeauthoryear{Pipino, Silk,
\& Matteucci}{2009}]{2009MNRAS.392..475P} Pipino A., Silk J., Matteucci F., 2009, MNRAS, 392, 475

\bibitem[\protect\citeauthoryear{Pontzen \&
    Governato}{2012}]{2012MNRAS.tmp.2641P} Pontzen A., Governato F., 2012,
    MNRAS, 2641

\bibitem[\protect\citeauthoryear{Pooley et al.}{2012}]{2012ApJ...744..111P}
Pooley D., Rappaport S., Blackburne J.~A., Schechter P.~L., Wambsganss J.,
2012, ApJ, 744, 111

%\bibitem[\protect\citeauthoryear{Prochaska \&
%    Hennawi}{2009}]{2009ApJ...690.1558P} Prochaska J.~X., Hennawi J.~F.,
%    2009, ApJ, 690, 1558

\bibitem[\protect\citeauthoryear{Ragone-Figueroa
\& Granato}{2011}]{2011MNRAS.414.3690R} Ragone-Figueroa C., Granato G.~L., 2011, MNRAS, 414, 3690

\bibitem[\protect\citeauthoryear{Read
\& Gilmore}{2005}]{2005MNRAS.356..107R} Read J.~I., Gilmore G., 2005, MNRAS, 356, 107

%\bibitem[]{}Renzini, A. 2006, ARA\&A, 44, 141

%\bibitem[]{}S{\'e}rsic, J.L. 1963, Boletin de la Asociacion Argentina de
%    Astronomia, 6, 41

\bibitem[\protect\citeauthoryear{Richtler et
al.}{2011}]{2011A&A...531A.119R} Richtler T., Salinas R., Misgeld I., Hilker M., Hau G.~K.~T., Romanowsky A.~J.,
Schuberth Y., Spolaor M., 2011, A\&A, 531, A119


\bibitem[\protect\citeauthoryear{Salucci
\& Frigerio Martins}{2009}]{2009EAS....36..133S} Salucci P., Frigerio Martins C., 2009, EAS, 36, 133

%\bibitem[\protect\citeauthoryear{Saracco, Longhetti, \&
%    Andreon}{2009}]{2009MNRAS.392..718S} Saracco P., Longhetti M., Andreon
%    S., 2009, MNRAS, 392, 718

%\bibitem[\protect\citeauthoryear{Saracco, Longhetti, \&
%    Gargiulo}{2010}]{2010MNRAS.408L..21S} Saracco P., Longhetti M.,
%    Gargiulo A., 2010, MNRAS, 408, L21

%\bibitem[]{}Shen, S., et al. 2003, MNRAS, 343, 978

\bibitem[\protect\citeauthoryear{Sijacki et
    al.}{2007}]{2007MNRAS.380..877S} Sijacki D., Springel V., Di Matteo
    T., Hernquist L., 2007, MNRAS, 380, 877

\bibitem[\protect\citeauthoryear{Silk
\& Rees}{1998}]{1998A&A...331L...1S} Silk J., Rees M.~J., 1998, A\&A, 331, L1
%\bibitem[]{}Somerville, R.S., et al. 2008, MNRAS, 391, 481

\bibitem[\protect\citeauthoryear{Somerville et
al.}{2008}]{2008MNRAS.391..481S} Somerville R.~S., Hopkins P.~F., Cox
T.~J., Robertson B.~E., Hernquist L., 2008, MNRAS, 391, 481

\bibitem[\protect\citeauthoryear{Sonnenfeld et
al.}{2011}]{2011arXiv1111.4215S} Sonnenfeld A., Treu T., Gavazzi R.,
Marshall P.~J., Auger M.~W., Suyu S.~H., Koopmans L.~V.~E., Bolton A.~S.,
2011, arXiv, arXiv:1111.4215

\bibitem[\protect\citeauthoryear{Springel}{2005}]{2005MNRAS.364.1105S}
    Springel V., 2005, MNRAS, 364, 1105

%\bibitem[]{}Springel, V., Di Matteo, T., \& Hernquist, L. 2005, MNRAS,
%    361, 776

%\bibitem[\protect\citeauthoryear{Ryan et al.}{2010}]{2010arXiv1007.1460R}
%    Ryan R.~E., Jr., et al., 2010, arXiv, arXiv:1007.1460
%\bibitem[]{}Tacconi, L.J., et al. 2008, ApJ, 680, 246


%\bibitem[]{}Tacconi, L.J., et al. 2010, Nature, 463, 781

%\bibitem[\protect\citeauthoryear{Toft et al.}{2007}]{2007ApJ...671..285T}
%    Toft S., et al., 2007, ApJ, 671, 285

\bibitem[\protect\citeauthoryear{Tonini, Lapi,
\& Salucci}{2006}]{2006ApJ...649..591T} Tonini C., Lapi A., Salucci P., 2006, ApJ, 649, 591

\bibitem[\protect\citeauthoryear{Tortora et
al.}{2012}]{2012arXiv1201.2945T} Tortora C., La Barbera F., Napolitano
N.~R., de Carvalho R.~R., Romanowsky A.~J., 2012, arXiv, arXiv:1201.2945

\bibitem[\protect\citeauthoryear{Treu et al.}{2010}]{2010ApJ...709.1195T}
Treu T., Auger M.~W., Koopmans L.~V.~E., Gavazzi R., Marshall P.~J., Bolton
A.~S., 2010, ApJ, 709, 1195


%\bibitem[]{}Trujillo, I., et al. 2004, ApJ, 604, 521


%\bibitem[\protect\citeauthoryear{Trujillo et
%    al.}{2006}]{2006ApJ...650...18T} Trujillo I., et al., 2006, ApJ, 650,
%    18

%\bibitem[]{}Trujillo, I., et al. 2007, MNRAS, 382, 109

%\bibitem[\protect\citeauthoryear{Trujillo, Ferreras, \& de la
%    Rosa}{2011}]{2011arXiv1102.3398T} Trujillo I., Ferreras I., de la Rosa
%    I.~G., 2011, arXiv, arXiv:1102.3398

%\bibitem[]{}Trujillo, I., et al. 2009, ApJ, 692, L118

%\bibitem[]{}van der Wel, A., et al. 2009, ApJ, 698, 1232

%\bibitem[\protect\citeauthoryear{Tutukov}{1978}]{1978A&A....70...57T}
%    Tutukov A.~V., 1978, A\&A, 70, 57

%\bibitem[\protect\citeauthoryear{Valentinuzzi et
%    al.}{2010}]{2010ApJ...712..226V} Valentinuzzi T., et al., 2010, ApJ,
%   712, 226

%\bibitem[]{}van der Wel, A., et al. 2008, ApJ, 688, 48

%\bibitem[]{}van Dokkum, P.G., et al. 2008, ApJ, 677, L5

%\bibitem[]{}van Dokkum, P.G., Kriek, M., \& Franx, M. 2009, Nature, 460,
%    717

%\bibitem[]{}van Dokkum, P.~G., et al. 2010, ApJ, 709, 1018

%\bibitem[\protect\citeauthoryear{Welch, Sage, \&
%    Young}{2010}]{2010ApJ...725..100W} Welch G.~A., Sage L.~J., Young
%    L.~M., 2010, ApJ, 725, 100

\bibitem[\protect\citeauthoryear{Viola et al.}{2008}]{2008MNRAS.383..777V}
    Viola M., Monaco P., Borgani S., Murante G., Tornatore L., 2008, MNRAS,
    383, 777

\bibitem{}Zhao, D.H., Mo, H.J., Jing, Y.P., \& B\"{o}rner, G.
    2003, MNRAS, 339, 12

\bibitem[\protect\citeauthoryear{Zitrin et al.}{2011}]{2011ApJ...742..117Z}
Zitrin A., et al., 2011, ApJ, 742, 117


%\bibitem[]{}Zirm, A.W., et al. 2007, ApJ, 656, 66

\end{thebibliography}
\end{document}